\documentclass[a4paper,11pt]{article}
\usepackage{pos}

\usepackage{xcolor}
\usepackage{multirow}
\usepackage{capt-of}
\usepackage{siunitx}
\usepackage{graphicx}
\usepackage{slashed}
\usepackage{tabularx}
\usepackage{multirow}
\usepackage{braket}
\usepackage{amsfonts}
\usepackage{mathtools}
\usepackage{mathrsfs}
\usepackage{bbold}
\usepackage{comment}
\usepackage[compat=1.1.0]{tikz-feynman}
\usepackage{subcaption}

\pgfmathsetmacro\sizedot{1.1}
\pgfmathsetmacro\sizesqdot{1.5}
\pgfmathsetmacro\sizecrodot{1.0}

\newcommand{\coeff}[2]{ \mathcal{C}_{#1} ^{#2} } 
\newcommand{\op}[2]{ \mathcal{O}_{#1} ^{#2} } 
\newcommand{\mtrx}[1]{\mathcal{M}_{#1}}
\newcommand{\order}[1] {\mathcal{O}\left( #1 \right)}

\newcommand{\highscale}{\Lambda}

\newcommand{\opdim}{\mathfrak{D}}

\newcommand{\lag}{ \mathcal{L}} %Lagrangian

\renewcommand{\logo}{\relax}

\title{ Higgs probes of top contact interactions and their interplay with Higgs self-coupling}
\author*{Stefano Di Noi}
\author{Ramona Gr\"ober}

\affiliation{Dipartimento di Fisica e Astronomia “G. Galilei”, Universit\`a di Padova,\\
  Via Marzolo 8,  Padua, Italy}

\affiliation{Istituto Nazionale di Fisica Nucleare, Sezione di Padova,\\
  Via Marzolo 8, Padua, Italy}

\emailAdd{stefano.dinoi@phd.unipd.it}
\emailAdd{ramona.groeber@pd.infn.it}

\abstract{
We present a method which relies on loop contributions from four-top SMEFT operators to single Higgs observables to contrain their Wilson coefficients. Such loop-induced terms have a non-trivial interplay with the extraction of the trilinear Higgs coupling. 
We show that this strategy	can, for some operators, lead to more stringent bounds than direct measurement via top quark data. Finally, we mention some recent developements in the treatment of $\gamma_5$ in dimensional regularisation in the context of the SMEFT.
}

\FullConference{The Eleventh Annual Conference on Large Hadron Collider Physics (LHCP2023)\\
 22-26 May 2023\\
 Belgrade, Serbia\\}

\begin{document}
\maketitle
\section{Introduction}
The Standard Model (SM) is one of the biggest scientific successes of our time. However, it leaves several phenomena unexplained. The next decades will not see a significant increase in the energy range of collider experiments, lowering the chances to directly observe new particles.

In the absence of New Physics (NP) signals, Effective Field Theories (EFTs) provide a powerful, pragmatic and general approach for the search of physics beyond the SM. The underlying strategy is to parametrise the effects of heavy new physics lying beyond our current experimental reach.
The Standard Model Effective Field Theory (SMEFT) is built adding to the SM Lagrangian a tower of higher-dimensional operators built with the SM fields, which must respect the SM gauge symmetry group $\mathbf{SU(3)}_{\mathrm{C}} \otimes\mathbf{SU(2)}_{\mathrm{L}}\otimes\mathbf{U(1)}_{\mathrm{Y}}$, 
\begin{equation}
	\label{eq:SMEFTlag}
	\lag_{\textrm{SMEFT}} =\lag_{\textrm{SM}} + \sum_{\opdim_i>4} \frac{\coeff{i}{}}{\highscale^{\opdim_i-4}} \op{i}{},
\end{equation}
being  $\opdim_i$ the mass dimension of $\op{i}{}$. This allows to describe the effects of heavy degrees of freedom with masses  of order $\Lambda \gtrsim v = 246\, \mathrm{GeV}$.  The $\coeff{i}{}$ are known as Wilson coefficients.
If we assume that lepton and baryon number are conserved, the first non trivial order of the SMEFT expansion leads to dimension-six operators. 
The inclusion of higher-dimensional operators entails many consequences: the interaction vertices are modified and new topologies arise (see \cite{Dedes:2017zog}) and the running of the SM parameters is modified (see \cite{rge1}). 

\section{Four-top operators and interplay with the trilinear Higgs self-coupling}
In many NP scenarios, new degrees of freedom are expected to have an enhanced coupling with particles of the third generation. This motivates us to focus on four-top operators:

\begin{equation} \label{eq:Lag4t}
\begin{split}
			\lag_{\opdim=6}^{4t} &= \frac{\coeff{tt}{}}{\highscale^2} (\bar{t}_R\gamma^\mu t_R)(\bar{t}_R\gamma_\mu t_R) + \frac{\coeff{QQ}{(1)}}{\highscale^2} (\bar{Q}_L\gamma^\mu  Q_L)(\bar{Q}_L\gamma_\mu Q_L) \\
			&+ \frac{\coeff{QQ}{(3)}}{\highscale^2} (\bar{Q}_L\gamma^\mu \tau^I Q_L)(\bar{Q}_L\gamma_\mu\tau^I Q_L) + \frac{\coeff{Qt}{(1)}}{\highscale^2} (\bar{Q}_L\gamma^\mu  Q_L)(\bar{t}_R\gamma_\mu t_R)+ \frac{\coeff{Qt}{(8)}}{\highscale^2} (\bar{Q}_L\gamma^\mu T^A Q_L)(\bar{t}_R\gamma_\mu T^A t_R) ,
		\end{split}
\end{equation}
where $Q_L$ and $t_R$ refer to the
third family iso-doublet (left-handed) and iso-singlet (right-handed, up-type) and $T^A$ ($\tau^I$) are the generators in the fundamental representation of $\mathbf{SU(3)}_{\mathrm{c}}$ ($\mathbf{SU(2)}_{\mathrm{L}}$).

A direct measurement of the four-top operators
requires the production of four top quarks, which is a rare process in the SM ($\sim 12$ fb, see \cite{Bevilacqua:2012em,Jezo:2021smh,Frederix:2017wme}).
Bounds obtained from a combination of Higgs, diboson and top quark data can be found in \cite{Ethier:2021bye}.  Some operators, such as $\op{Qt}{(1)}$, can be constrained via electroweak precision data, see \cite{Dawson:2022bxd}.  Moreover, also flavour observables can provide bounds on flavour-conserving operators as the ones discussed here, see \cite{Silvestrini:2018dos}.

\begin{figure}[h]
	\centering
	\begin{subfigure}[t]{0.3\textwidth}
		\centering
		\begin{tikzpicture} 
			\begin{feynman}[small]
				\vertex  (g1)  {$g$};
				\vertex  (gtt1) [dot,scale=\sizedot,right = of g1] {};
				\vertex (4F) [square dot, scale=\sizesqdot,below right= of gtt1,color=red] {};
				\vertex  (gtt2) [dot,scale=\sizedot,below left = of 4F] {};
				\vertex  (g2) [left=of gtt2]  {$g$};
				\vertex (htt) [dot,scale=\sizedot,right = 20 pt  of 4F] {};
				\vertex (h) [right = of htt] {$h$};
				
				\diagram* {
					(g1)  -- [gluon] (gtt1),
					(g2) -- [gluon] (gtt2),
					(h)  -- [scalar] (htt),
					(gtt1) -- [fermion] (4F) 
					-- [fermion] (gtt2)
					-- [fermion] (gtt1), 
					(4F) -- [fermion, half left] (htt) -- [fermion, half left] (4F)
				};
			\end{feynman}
		\end{tikzpicture}
		\caption{Contribution to the Higgs-top quark coupling.}\label{fig:4t_2loop_yt}
	\end{subfigure}
	\begin{subfigure}[t]{0.3 \textwidth}
		\centering
		\begin{tikzpicture} 
			\begin{feynman}[small]
				\vertex  (g1)  {$g$};
				\vertex  (gtt1) [dot,scale=\sizedot,right = of g1] {};
				\vertex (htt) [dot,scale=\sizedot,below right = of gtt1] {};
				\vertex (4F) [square dot, scale = \sizesqdot,color=red,below right= 14 pt of gtt1] {};
				\vertex (inv1) [scale=0.01,above right = 18 pt of 4F] {};
				\vertex  (gtt2) [dot,scale=\sizedot,below left = of htt] {};
				\vertex  (g2) [left=of gtt2]  {$g$};
				\vertex (h) [right = of htt] {$h$};
				\diagram* {
					(g1)  -- [gluon] (gtt1),
					(g2) -- [gluon] (gtt2),
					(h)  -- [scalar] (htt),
					(gtt1) -- [fermion] (4F) -- [fermion] (htt) 
					-- [fermion] (gtt2)
					-- [fermion] (gtt1), 
					(4F) -- [fermion, half left] (inv1) -- [fermion, half left] (4F)
				};
			\end{feynman}
		\end{tikzpicture}
		\caption{Contribution to the top quark propagator.}\label{fig:4t_2loop_mt}
	\end{subfigure}
	\begin{subfigure}[t]{0.3 \textwidth}
		\centering
		\begin{tikzpicture} 
			\begin{feynman}[small]
				\vertex  (g1)  {$g$};
				\vertex (gtt1) [dot,scale=\sizedot,right= 28 pt of g1] {};
				\vertex  (4F) [square dot, scale=\sizesqdot, right = 20 pt of gtt1,color=red] {};
				\vertex (htt) [dot,scale=\sizedot,below right = of 4F] {};
				\vertex  (gtt2) [dot,scale=\sizedot,below left = of htt] {};
				\vertex  (g2) [left= 50 pt of gtt2]  {$g$};
				\vertex (h) [right = of htt] {$h$};
				
				\diagram* {
					(g1)  -- [gluon] (gtt1),
					(g2) -- [gluon] (gtt2),
					(h)  -- [scalar] (htt),
					(gtt1) -- [fermion, half right] (4F) 
					-- [fermion,half right] (gtt1),
					(4F) -- [fermion] (htt) -- [fermion] (gtt2) -- [fermion] (4F)
				};
			\end{feynman}
		\end{tikzpicture}
		\caption{Contribution to the gluon-top quark vertex.}\label{fig:4t_2loop_gtt}
	\end{subfigure}
	\caption{Contributions from insertions of four-top operators (red square dot) to $gg \to h$ at two-loop level. Black dots denote SM interactions.} \label{fig:4t_2loop}
\end{figure}
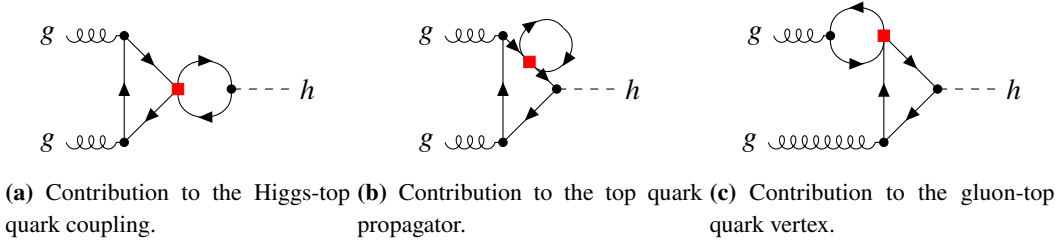

Here we discuss a method to indirectly constrain four-top operators via single Higgs observables where they enter at loop level, first presented in \cite{Alasfar:2022zyr}, such as $gg \to h$, $h \to gg/\gamma \gamma$ and $\bar{t}th$ production. The process we showcase here is $gg \to h$, which receives contributions from the operators in Eq.~\eqref{eq:Lag4t} at two-loop level (see Fig.~\ref{fig:4t_2loop}).

\begin{figure}[h]
\begin{subfigure}[t]{0.3 \textwidth}
	\centering
\begin{tikzpicture}[baseline=(l3)]
	\begin{feynman}[small]
		\vertex (g1) {\(  g \)};
		\vertex (gs1) [dot, scale=\sizedot, right of=g1] {};
		\vertex (yt1) [dot, scale=\sizedot, right of=gs1] {};
		\vertex (l3) [blue, square dot, scale=\sizesqdot, below right = 30 pt of yt1] {};	
		\vertex (yt2) [dot, scale=\sizedot,below left= 30 pt of l3] {};
		\vertex (gs2) [dot, scale=\sizedot, left of=yt2] {};
		\vertex (g2) [left of= gs2] {\(  g \)};
		\vertex (hf) [right = of l3] {\( h \)};
		\diagram* {
			(g1)  -- [gluon] (gs1),
			(g2)  -- [gluon] (gs2),
			(gs1) -- [fermion] (yt1) --[fermion] (yt2) -- [fermion] (gs2) -- [fermion] (gs1),
			(yt1) -- [scalar] (l3) -- [scalar] (yt2),
			(l3) -- [scalar] (hf)
		};
\end{feynman}  \end{tikzpicture} 
\subcaption{}  \label{fig:lambda_2loop_1}
\end{subfigure} %
\begin{subfigure}[t]{0.3 \textwidth}
	\centering 
\begin{tikzpicture}[baseline=(l3)]
	\begin{feynman}[small]
		\vertex (g1) {\(  g \)};
		\vertex (gs1) [dot, scale=\sizedot, below right =  of g1] {};
		\vertex (g2) [below left = of gs1] {\(  g \)};
		\vertex (gs2) [dot, scale=\sizedot,right = 20 pt of gs1] {};
		\vertex (yt1) [dot, scale=\sizedot, above = 20 pt of gs2] {};
		\vertex (yt2) [dot, scale=\sizedot, below =  20 pt of gs2] {};
		\vertex (l3) [blue, square dot, scale=\sizesqdot, right  = 20 pt of gs2] {};	
		\vertex (hf) [right = of l3] {\( h \)};
		\diagram* {
			(yt1) -- [quarter left, scalar] (l3) -- [quarter left, scalar] (yt2),
			(l3) -- [scalar] (hf),
			(gs1)-- [fermion, quarter left] (yt1) -- [fermion] (gs2) -- [fermion] (yt2) --[fermion, quarter left] (gs1),
			(g1) --[gluon] (gs1),
			(g2) --[gluon] (gs2),						
		};
	\end{feynman}  
\end{tikzpicture} 
\subcaption{} \label{fig:lambda_2loop_2}
\end{subfigure} %
\begin{subfigure}[t]{0.3 \textwidth}
	\centering
	\begin{tikzpicture}[baseline=(l3)]
		\begin{feynman}[small]
			\vertex (g1) {\(  g \)};
			\vertex (gs1) [dot, scale=\sizedot, right of=g1] {};
			\vertex (yt1) [dot, scale=\sizedot, below right of=gs1] {};
			\vertex (l31) [dot, scale=\sizesqdot,  right =15 pt of  yt1] {};	
			\vertex (l32) [blue, square dot, scale=\sizesqdot,  right =20 pt of  l31] {};	
			\vertex (gs2) [dot, scale=\sizedot, below left of= yt1] {};
			\vertex (g2) [left of= gs2] {\(  g \)};
			\vertex (hf) [right = 20 pt of l32] {\( h \)};
			\diagram* {
				(g1)  -- [gluon] (gs1),
				(g2)  -- [gluon] (gs2),
				(gs1) -- [fermion] (yt1) -- [fermion] (gs2) -- [fermion] (gs1),
				(yt1) -- [scalar] (l31) ,
				(l31)-- [scalar, half right] (l32) -- [scalar, half right] (l31),
				(l32) -- [scalar] (hf)
			};
	\end{feynman}  \end{tikzpicture} 
	\subcaption{} \label{fig:lambda_2loop_univ}
\end{subfigure} %
	\caption{Contributions from insertions of a modified trilinear coupling (blue square dot) to $gg \to h$ at two-loop level. Black dots denote SM interactions.} \label{fig:lambda_2loop} 
\end{figure}
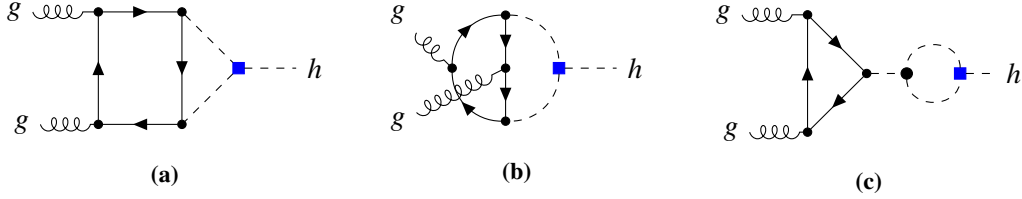

In the SMEFT the modifications of the trilinear Higgs coupling are parametrised by $\op{\phi}{} \equiv  \left( \phi^\dagger \phi \right)^3$. The trilinear coupling impacts $gg \to h$ at two-loop order, see Fig.~\ref{fig:lambda_2loop}. Some contributions are process-specific and linear in the trilinear coupling (as the ones in Figs.~\ref{fig:lambda_2loop_1}, \ref{fig:lambda_2loop_2}) while others are universal and quadratic in the trilinear coupling (as in Fig.~\ref{fig:lambda_2loop_univ}), see \cite{Degrassi:2016wml,Gorbahn:2016uoy} for the details.

The bare matrix element in our case is (using dimensional regularisation with $D=4-2\epsilon$), schematically, $ i \mtrx{\mathrm{bare}} = A \left( \frac{1}{\epsilon} + \log \frac{\mu_R^2}{\highscale^2}  \right) + B  $. To get rid of the divergent part, a renormalisation procedure is required	and we have $ i \mtrx{\mathrm{ren}}= i \mtrx{\mathrm{bare}}+ i \mtrx{\mathrm{c.t.}} = A  \log \frac{\mu_R^2}{\highscale^2} + B  $. The logarithmic term is connected to the divergences of the theory, thus to the anomalous dimension which, for the SMEFT at one-loop level, is available in \cite{rge1,rge2,rge3}. It follows that these terms can be inferred from the anomalous dimension and do not require any computation. Conversely, the finite terms, represented by $B$, require a full computation. As it was shown in \cite{Alasfar:2022zyr}, such contributions can be comparable with the logarithmic-enhanced ones and thus be phenomenologically relevant.

 Several subleties arise due to the delicate question of the continuation of $\gamma_5$, appearing in the four-top vertices, to $D\ne 4$ dimensions. We performed the computation using the two most common continuation schemes, namely NDR (\cite{CHANOWITZ1979225}) and BMHV (\cite{THOOFT1972189,Breitenlohner:1977hr}). We find a divergence only in the first case,yielding an anomalous dimension for $\op{\phi G}{}= \left(\phi^\dagger \phi \right) G_{\mu\nu}^A G^{\mu\nu,A}$ which depends on the scheme. This phenomenon is well-known in the context of the $\order{\alpha_s}$ contributions to transitions as $b \to s \gamma/ s g$ (see \cite{Ciuchini:1993ks,Ciuchini:1993fk}) and it is due to a non-trivial interplay between different effective operators. We presented in \cite{DiNoi:2023ygk} an extension of this strategy, which allows to obtain a scheme-independent result not only for the anomalous dimension but also for the (potentially relevant) finite parts. This can be achieved assuming that the Wilson coefficients depend on the scheme chosen for $\gamma_5$. Moreover, we presented in \cite{DiNoi:2023ygk} an explicit map to translate the Wilson coefficients from one $\gamma_5$ scheme to the other.

\begin{figure}[]
	\centering
	\begin{subfigure}[t]{.8\linewidth}
		\centering
		\includegraphics[width=.7\linewidth]{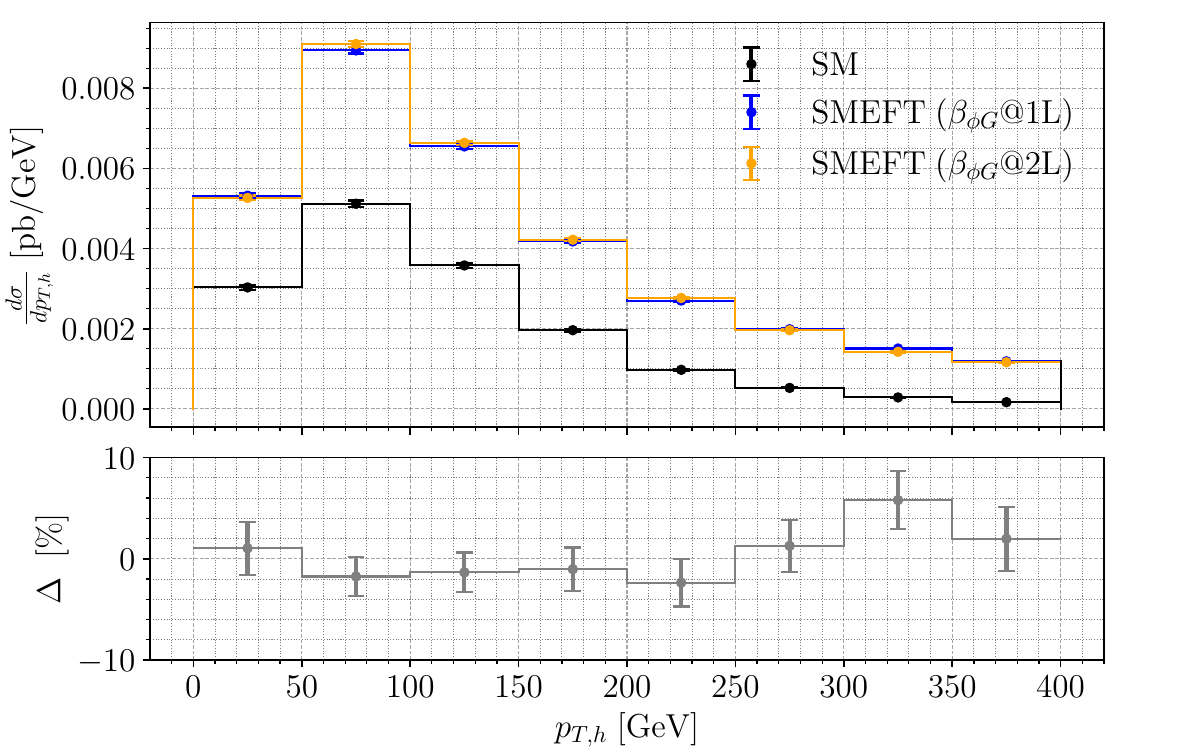}
		\caption{Higgs transverse momentum distribution (upper panel) and percentual difference for each bin between the one- and two-loop running computed as $\Delta \equiv  \left({
				\left( \frac{d \sigma }{d p_{T,h}} \right)_{\mathrm{1L}}-\left( \frac{d \sigma }{d p_{T,h}} \right)_{\mathrm{2L}}
			} \right)/{\left( \frac{d \sigma }{d p_{T,h}} \right)_{\mathrm{2L}}
			}$ (lower panel).}
		\label{fig:dsigma1Lvs2L}
	\end{subfigure}
	\begin{subfigure}[t]{.8\linewidth}
		\centering 
		\includegraphics[width=.7 \linewidth]{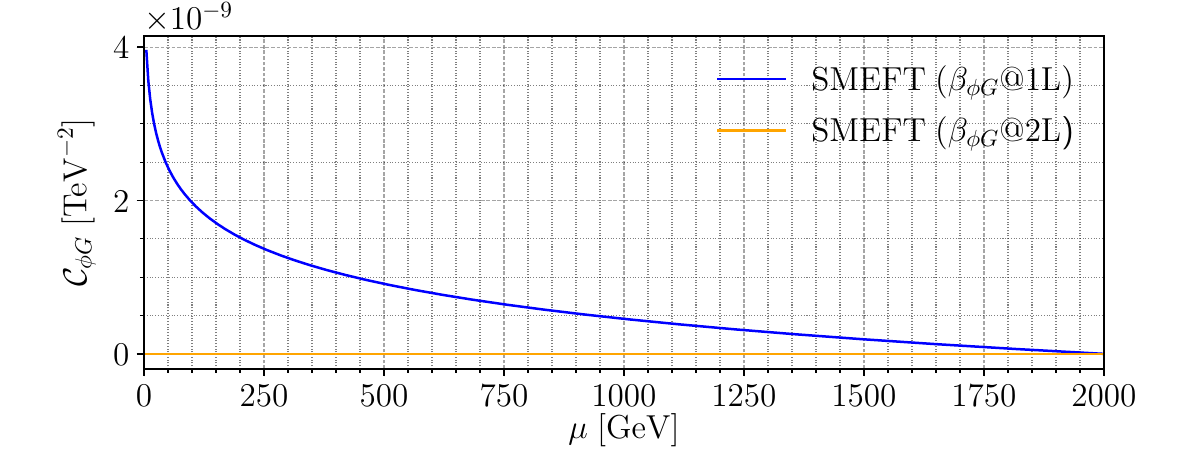}
		\caption{Running of $\coeff{\phi G}{}$.}
		\label{fig:CHGRunning}
	\end{subfigure}
	\caption{Comparison between the one- and two-loop running of $\coeff{\phi G}{}$.}
	\label{fig:1Lvs2L}
\end{figure}
The effect of the scheme-dependent anomalous dimension is discussed in Fig.~\ref{fig:1Lvs2L} for $pp \to \bar{t}t h$.\footnote{A detailed study of the renormalisation group effects for $pp \to \bar{t}t h$ can be found in \cite{DiNoi:2023onw}.}
The two cases do not exhibit a large difference in this scenario, being a two-loop effect while the SM amplitude arises at tree-level. In other processes, scheme-dependent running effects can be more relevant and require a coherent treatment. 
A detailed study which takes into account also the scheme-dependence of the finite parts, extending the results presented in \cite{Alasfar:2022zyr}, is left for future work. 
\section{Conclusions and outlook}
In this talk, we presented an indirect method to constrain SMEFT four-top operators using their loop-induced contribution to single Higgs observables. The bounds obtained with this strategy offer an alternative way to direct probes, yielding in some cases an even better result (see \cite{Alasfar:2022zyr}).

The promising program of indirect bounds on four-top operators can be extended by considering differential observables to increase the constraining power of the fit. Moreover, including differential observables may remove some degeneracies in the parameter space, see \cite{DiVita:2017eyz}.
A possible extension of this analysis is $pp \to hj$, which we leave for future work.
The impact of four-top operators in $pp \to hh$ (considering also the subleties of the continuation of $\gamma_5$) has been studied in \cite{Heinrich:2023rsd}.

\bibliographystyle{unsrt} % Or use \bibliographystyle{plain}

\begin{thebibliography}{99}
%	\bibitem{dim6smeft}
%	B. Grzadkowski, M. Iskrzynski, M. Misiak, J. Rosiek,
%	\textit{Dimension-Six Terms in the Standard Model Lagrangian},
%	JHEP \textbf{10}, 085 (2010),
%	\href{https://doi.org/10.1007/JHEP10(2010)085}{doi:10.1007/JHEP10(2010)085},arXiv:1008.4884 [hep-ph].
	
	\bibitem{Dedes:2017zog}
	A. Dedes, W. Materkowska, M. Paraskevas, J. Rosiek, K. Suxho,
	\textit{Feynman rules for the Standard Model Effective Field Theory in $R_{\xi}$-gauges},
	JHEP \textbf{06}, 143 (2017),
	\href{https://doi.org/10.1007/JHEP06(2017)143}{doi:10.1007/JHEP06(2017)143},
	arXiv:1704.03888 [hep-ph].
	
	\bibitem{rge1}
	Elizabeth E. Jenkins, Aneesh V. Manohar, Michael Trott,
	\textit{Renormalization Group Evolution of the Standard Model Dimension Six Operators I: Formalism and $\lambda$ Dependence},
	JHEP \textbf{10}, 087 (2013),
	\href{https://doi.org/10.1007/JHEP10(2013)087}{doi:10.1007/JHEP10(2013)087},
	arXiv:1308.2627 [hep-ph].
	
	
	\bibitem{Bevilacqua:2012em}
	G. Bevilacqua and M. Worek,
	\textit{Constraining BSM Physics at the LHC: Four top final states with NLO accuracy in perturbative QCD},
	JHEP \textbf{07}, 111 (2012),
	\href{https://doi.org/10.1007/JHEP07(2012)111}{doi:10.1007/JHEP07(2012)111},
	arXiv:1206.3064 [hep-ph].
	
	\bibitem{Jezo:2021smh}
	Tom\'a\v{s} Je\v{z}o and Manfred Kraus,
	\textit{Hadroproduction of four top quarks in the powheg box},
	Phys. Rev. D \textbf{105} (11), 114024 (2022),
	\href{https://doi.org/10.1103/PhysRevD.105.114024}{doi:10.1103/PhysRevD.105.114024},
	arXiv:2110.15159 [hep-ph].
	
	\bibitem{Frederix:2017wme}
	Rikkert Frederix, Davide Pagani, Marco Zaro,
	\textit{Large NLO corrections in \(t\bar{t}W^{\pm}\) and \(t\bar{t}t\bar{t}\) hadroproduction from supposedly subleading EW contributions},
	JHEP \textbf{02}, 031 (2018),
	\href{https://doi.org/10.1007/JHEP02(2018)031}{doi:10.1007/JHEP02(2018)031},
	arXiv:1711.02116 [hep-ph].
	
	
	\bibitem{Ethier:2021bye}
	Jacob J. Ethier, Giacomo Magni, Fabio Maltoni, Luca Mantani, Emanuele R. Nocera, Juan Rojo, Emma Slade, Eleni Vryonidou, Cen Zhang; SMEFiT Collaboration,
	\textit{Combined SMEFT interpretation of Higgs, diboson, and top quark data from the LHC},
	JHEP \textbf{11}, 089 (2021),
	\href{https://doi.org/10.1007/JHEP11(2021)089}{doi:10.1007/JHEP11(2021)089},
	arXiv:2105.00006 [hep-ph].
	
	
	\bibitem{Dawson:2022bxd}
	Sally Dawson, Pier Paolo Giardino,
	\textit{Flavorful electroweak precision observables in the Standard Model effective field theory},
	Phys. Rev. D \textbf{105}, 073006 (2022),
	\href{https://doi.org/10.1103/PhysRevD.105.073006}{doi:10.1103/PhysRevD.105.073006},
	arXiv:2201.09887 [hep-ph].

	
	\bibitem{Silvestrini:2018dos}
	Luca Silvestrini, Mauro Valli,
	\textit{Model-independent Bounds on the Standard Model Effective Theory from Flavour Physics},
	Phys. Lett. B \textbf{799}, 135062 (2019),
	\href{https://doi.org/10.1016/j.physletb.2019.135062}{doi:10.1016/j.physletb.2019.135062},
	arXiv:1812.10913 [hep-ph].	
	
		\bibitem{Alasfar:2022zyr}
	Lina Alasfar, Jorge de Blas, Ramona Gr\"ober,
	\textit{Higgs probes of top quark contact interactions and their interplay with the Higgs self-coupling},
	JHEP \textbf{05}, 111 (2022),
	\href{https://doi.org/10.1007/JHEP05(2022)111}{doi:10.1007/JHEP05(2022)111},
	arXiv:2202.02333 [hep-ph].
	
	\bibitem{Degrassi:2016wml}
	Giuseppe Degrassi, Pier Paolo Giardino, Fabio Maltoni, Davide Pagani,
	\textit{Probing the Higgs self coupling via single Higgs production at the LHC},
	JHEP \textbf{12}, 080 (2016),
	\href{https://doi.org/10.1007/JHEP12(2016)080}{doi:10.1007/JHEP12(2016)080},
	arXiv:1607.04251 [hep-ph].
	
	\bibitem{Gorbahn:2016uoy}
	Martin Gorbahn, Ulrich Haisch,
	\textit{Indirect probes of the trilinear Higgs coupling: $gg \to h$ and $h \to \gamma \gamma$},
	JHEP \textbf{10}, 094 (2016),
	\href{https://doi.org/10.1007/JHEP10(2016)094}{doi:10.1007/JHEP10(2016)094},
	arXiv:1607.03773 [hep-ph].
	
	
	\bibitem{rge2}
	Elizabeth E. Jenkins, Aneesh V. Manohar, Michael Trott,
	\textit{Renormalization Group Evolution of the Standard Model Dimension Six Operators II: Yukawa Dependence},
	JHEP \textbf{01}, 035 (2014),
	\href{https://doi.org/10.1007/JHEP01(2014)035}{doi:10.1007/JHEP01(2014)035},
	arXiv:1310.4838 [hep-ph].
	
	\bibitem{rge3}
	Rodrigo Alonso, Elizabeth E. Jenkins, Aneesh V. Manohar, Michael Trott,
	\textit{Renormalization Group Evolution of the Standard Model Dimension Six Operators III: Gauge Coupling Dependence and Phenomenology},
	JHEP \textbf{04}, 159 (2014),
	\href{https://doi.org/10.1007/JHEP04(2014)159}{doi:10.1007/JHEP04(2014)159},
	arXiv:1312.2014 [hep-ph].
	
 \bibitem{CHANOWITZ1979225}
M. Chanowitz, M. Furman, I. Hinchliffe,
\textit{The axial current in dimensional regularization},
Nuclear Physics B \textbf{159} (1), 225--243 (1979),
\href{https://doi.org/10.1016/0550-3213(79)90333-X}{doi:10.1016/0550-3213(79)90333-X},
\url{https://www.sciencedirect.com/science/article/pii/055032137990333X}.

	\bibitem{THOOFT1972189}
	G. 't Hooft and M. Veltman,
	\textit{Regularization and renormalization of gauge fields},
	Nuclear Physics B \textbf{44} (1), 189--213 (1972),
	\href{https://doi.org/10.1016/0550-3213(72)90279-9}{doi:10.1016/0550-3213(72)90279-9},
	\url{https://www.sciencedirect.com/science/article/pii/0550321372902799}.
	
	 \bibitem{Breitenlohner:1977hr}
	P. Breitenlohner and D. Maison,
	\textit{Dimensional Renormalization and the Action Principle},
	Commun. Math. Phys. \textbf{52}, 11--38 (1977),
	\href{https://doi.org/10.1007/BF01609069}{doi:10.1007/BF01609069}.

	
	\bibitem{Ciuchini:1993ks}
	Marco Ciuchini, E. Franco, G. Martinelli, L. Reina, L. Silvestrini,
	\textit{Scheme independence of the effective Hamiltonian for \(b \rightarrow s \gamma\) and \(b \rightarrow s g\) decays},
	Phys. Lett. B \textbf{316}, 127--136 (1993),
	\href{https://doi.org/10.1016/0370-2693(93)90668-8}{doi:10.1016/0370-2693(93)90668-8},
	arXiv:hep-ph/9307364.
	
	
	\bibitem{Ciuchini:1993fk}
	Marco Ciuchini, E. Franco, L. Reina, L. Silvestrini,
	\textit{Leading order QCD corrections to \(b \rightarrow s \gamma\) and \(b \rightarrow s g\) decays in three regularization schemes},
	Nucl. Phys. B \textbf{421}, 41--64 (1994),
	\href{https://doi.org/10.1016/0550-3213(94)90223-2}{doi:10.1016/0550-3213(94)90223-2},
	arXiv:hep-ph/9311357.
	
	
	\bibitem{DiNoi:2023ygk}
	Stefano Di Noi, Ramona Gr\"ober, Gudrun Heinrich, Jannis Lang, Marco Vitti,
	\textit{On $\gamma_5$ schemes and the interplay of SMEFT operators in the Higgs-gluon coupling},
	arXiv:2310.18221 [hep-ph], October 2023.
	
	 \bibitem{DiNoi:2023onw}
	Stefano Di Noi and Ramona Gr\"ober,
	\textit{Renormalisation group running effects in \(pp\rightarrow t\bar{t}h\) in the Standard Model Effective Field Theory},
	arXiv:2312.11327 [hep-ph] (December 2023).
	
	 \bibitem{DiVita:2017eyz}
	Stefano Di Vita, Christophe Grojean, Giuliano Panico, Marc Riembau, Thibaud Vantalon,
	\textit{A global view on the Higgs self-coupling},
	JHEP \textbf{09}, 069 (2017),
	\href{https://doi.org/10.1007/JHEP09(2017)069}{doi:10.1007/JHEP09(2017)069},
	arXiv:1704.01953 [hep-ph].
	
 \bibitem{Heinrich:2023rsd}
Gudrun Heinrich, Jannis Lang,
\textit{Combining chromomagnetic and four-fermion operators with leading SMEFT operators for $gg\to hh$ at NLO QCD},
arXiv:2311.15004 [hep-ph], November 2023.
	

	
	

	
	






	



\end{thebibliography}

\end{document}